\documentclass[12pt]{iopart}

\usepackage{iopams}
\usepackage{mathrsfs}
\usepackage{graphicx}

\usepackage{cite}

\usepackage{color}

\def\D{\mathrm{D}}

\def\I{\mathscr{I}}
\def\O{\mathscr{O}}

\def\d{\mathrm{d}}
\def\h{\mathrm{h}}
\def\s{\mathrm{s}}
\def\u{\mathrm{u}}

\def\Rset{\mathbb{R}}

\def\id{\mathrm{id}}

\def\sech{\,\mathrm{sech}}

\def\epsilon{\varepsilon}

\renewcommand{\topfraction}1
\renewcommand{\bottomfraction}1

\begin{document}

\title[]%
{Nonintegrability of an extensible conducting rod in a uniform magnetic field}

\author{G.H.M. van der Heijden$^1$  and K.~Yagasaki$^2$}

\address{$^1$ Centre for Nonlinear Dynamics and its Applications,
Department of Civil, Environmental and Geomatic Engineering,
University College London,
Gower Street, London WC1E 6BT, UK}

\address{$^2$ Mathematics Division,
Department of Information Engineering,
Niigata Univeristy,
Ikarashi 2-no-cho, Nishi-ku, Niigata 950-2181, Japan}
\ead{\mailto{g.heijden@ucl.ac.uk}, \mailto{yagasaki@ie.niigata-u.ac.jp}}

\begin{abstract}
The equilibrium equations for an isotropic Kirchhoff rod are known to
be completely integrable. It is also known that neither the effects
of extensibility and shearability nor the effects of a uniform magnetic
field individually break integrability. Here we show, by means of a
Melnikov-type analysis, that, when combined, these effects do break
integrability giving rise to spatially chaotic configurations of the rod.
A previous analysis of the problem suffered from the presence of an
Euler-angle singularity. Our analysis provides an example of how in a
system with such a singularity a Melnikov-type technique can be applied
by introducing an artificial unfolding parameter. This technique can
be applied to more general problems.
\end{abstract}

\pacs{46.25.Hf, 02.30.Ik, 05.45.Ac}
\maketitle

\section{Introduction}

The geometrically exact static equilibrium equations for a uniform
symmetric (i.e., transversely isotropic) elastic rod are well known
to be completely integrable \cite{KM97}. In fact,
there is a close relationship between these equations and those describing
the dynamics of spinning tops \cite{L27}. It is also known
that some perturbations of the rod equations are integrable, but that
others are not. For instance, anisotropy of the cross-section \cite{MH88}
and intrinsic curvature \cite{CHT97} destroy integrability, as does the
effect of gravity \cite{GP99}, but extensibility and shearability \cite{S00}
do not, nor does the effect of an external force due to a uniform magnetic
field \cite{SH08}.

In this paper we show by means of a Melnikov-type analysis that although
extensibility and magnetic field individually do not destroy integrability,
their combined effect does. The results may be relevant for the study of
(localised) spatial configurations of electrodynamic space tethers, i.e.,
conducting cables that exploit the earth's magnetic field to generate thrust
and drag (Lorentz) forces for manoeuvring \cite{CM08,VH10}.

The problem was studied by one of us in a previous paper \cite{SH09}, where
a two-degrees-of-freedom system was obtained by reducing the rod equations in
Lie-Poisson form to a canonical Hamiltonian system in terms of Euler angles
and their canonical momenta. Numerical evidence of spatially complex rod
configurations was given in the form of chaotic Poincar\'e sections. A
Melnikov analysis was also attempted but this was inconclusive as the version
of Melnikov theory used, the Hamiltonian extension to two-degrees-of-freedom
systems by Holmes \& Marsden \cite{HM83}, was not valid due to the polar
singularity associated with Euler angles. This singularity prevents the
partial differentiability of the Hamiltonian with respect to the action
variable that is required in the symmetry reduction from the
two-degrees-of-freedom Hamiltonian system to a periodically perturbed
planar system where the classical Melnikov result \cite{M63,GH83} can be
applied.

One way around the singularity would be to use a different set of coordinates;
for instance the Deprit-Andoyer coordinates used in several of the other
studies \cite{MH88,LK04}. Here we use a different method that
 removes the singularity from the unperturbed equilibrium with homoclinic
 orbits by introducing an artificial unfolding that corresponds to a change
 of coordinates.
By a careful scaling of the equations, the magnetic rod is regarded
 as the unperturbed problem and extensibility/shearability is viewed
 as the perturbation (the opposite to the scaling used in \cite{SH09}).
We then apply a version of Melnikov's method due to Lerman and Umanskii
\cite{LU84} and Wiggins \cite{W88}, briefly reviewed in the Appendix for
the reader's convenience, to show that there exist transverse homoclinic
orbits to a saddle-focus.
The existence of such orbits implies via a classical result
 of Devaney \cite{D76} that the equilibrium equations for an extensible
 conducting rod are nonintegrable.

Apart from its physical interest our result is also of interest because it
gives an example of how a Melnikov analysis can be applied in the presence
of an Euler-angle singularity. Such singularities arise naturally in many
mechanical problems, especially in canonically reduced Lie-Poisson systems,
and often coincide with the saddle-type solution to which homoclinic orbits
are sought. In our Melnikov application the singularity is circumvented by
introducing an artificial unfolding
that does not affect the physics of the problem.

This paper is organised as follows. In Section 2 we briefly review the
canonically reduced Hamiltonian derived in \cite{SH08} and introduce the
nondimensionalisation that will be used in subsequent sections. In Section 3
we introduce our scaling of the problem and identify the unperturbed system
and its homoclinic orbit as well as the perturbation. The Melnikov analysis
is then carried out in Section 4 and numerical illustration of the analytical
 result 
is given in Section 5.
Finally, a summary and some comments are given in Section 6.

\section{Hamiltonian for an extensible rod in a uniform magnetic field}

The equilibrium equations for an extensible and shearable rod in a uniform
magnetic field were derived in \cite{SH09}. The equations were obtained by
reduction of a nine-dimensional system in Lie-Poisson form with three
Casimirs to a six-dimensional canonical Hamiltonian system in terms of three
Euler angles $(\theta,\psi,\phi)$ and their canonical momenta
$(p_\theta,p_\psi,p_\phi)$, similar to the reduction of the isotropic elastic
rod \cite{HT00} or the symmetric heavy top \cite{G80}. In the case of a
transversely isotropic rod (i.e., one having equal bending and shear
stiffnesses in the two principal directions of the rod's cross-section)
the Hamiltonian is given by
\begin{eqnarray}
\fl
H\left( \theta, \psi, \phi, p_{\theta}, p_{\psi}, p_{\phi} \right) &
= \frac{1}{2 B} p_{\theta}^{2} + \frac{1}{2 B} \left( \frac{p_{\psi}-p_{\phi}
\cos\theta}{\sin\theta} \right)^{2} + C_{2}\cos\theta\left( \frac{C_{2}}{2}
\left(\frac{1}{K}-\frac{1}{J}\right)\cos\theta + 1 \right) \nonumber \\
\fl
& \hspace{0.75cm} + \left( C_{2}\left(\frac{1}{K}-\frac{1}{J}\right)\cos\theta
+ 1 \right)\sin\theta\cos\psi\sqrt{ 2 C_{1}^{} - C_{2}^{2} - 2 \lambda
p_{\psi} } \nonumber \\ 
\fl
& \hspace{1.25cm} + \frac{1}{2}\left(\frac{1}{K}-\frac{1}{J}\right)\sin^{2}
\theta\cos^{2}\psi\left( 2 C_{1}^{} - C_{2}^{2} - 2 \lambda p_{\psi} \right)
-\frac{\lambda}{J}p_\psi. \label{eq:full_ham}
\end{eqnarray}
Here $B$ is the bending stiffness, $J$ the shear stiffness, $K$ the axial
stiffness and $\lambda=IB_m$ the magnetic parameter with $I$ the current
in the rod and $B_m$ the strength of the magnetic field, assumed uniform.
$C_1$ and $C_2$ are constant values of two of the Casimir first integrals
of the system. $C_1$ is related to the magnitude of force in the system,
while $C_2$ is the force in the direction of the magnetic field. These
constants are typically fixed by the boundary conditions.

Note that $\phi$ is a cyclic variable and hence $p_\phi$ (the twisting moment
in the rod) a constant. If, in addition, either the rod is inextensible and
unshearable ($1/J=1/K=0$) or the magnetic field (or electric current) is zero
($\lambda=0$) then Hamilton's equations corresponding to \eref{eq:full_ham}
have the first integral
\begin{eqnarray}
\fl
F & = C_{2} p_{\psi} + \lambda B \cos\theta - \sqrt{ 2 C_{1}^{} -
C_{2}^{2} - 2 \lambda p_{\psi}^{} } \left( p_{\theta}\sin\psi - \cos\psi
\left( \frac{p_{\phi} - p_{\psi}\cos\theta}{\sin\theta} \right) \right),
\label{eq:constraint}
\end{eqnarray}
rendering the system completely integrable \cite{SH08}.

We use the constants $C_2$ and $p_\phi$ to introduce dimensionless quantities
by setting
\begin{eqnarray*}
&& \bar{p}_\theta = \frac{p_\theta}{p_\phi}, \quad
\bar{p}_\psi = \frac{p_\psi}{p_\phi}, \quad \alpha = \frac{BC_2}{p_\phi^2},
\quad \gamma = C_2\left(\frac{1}{K}-\frac{1}{J}\right), \\
&& \mu=\frac{B^2(2C_1-C_2^2)}{p_\phi^4}, \quad \nu=\frac{\lambda B^2}{p_\phi^3},
\quad \epsilon = \frac{\lambda B}{Jp_\phi},
\end{eqnarray*}
so that the dimensionless Hamiltonian $\bar{H}=HB/p_\phi^2$ and integral
$\bar{F}=F/(C_2p_\phi)$ become
\begin{eqnarray}
\bar{H}\left( \theta, \psi, \bar{p}_{\theta}, \bar{p}_{\psi}\right)
 = & \frac{1}{2} \bar{p}_{\theta}^{2} + \frac{1}{2}
\left( \frac{\bar{p}_{\psi}-\cos\theta}{\sin\theta} \right)^{2} +
\alpha\cos\theta\left(1 + \frac{\gamma}{2}\cos\theta\right) \nonumber \\
& + (1+\gamma\cos\theta)\sin\theta\cos\psi\sqrt{\mu-2\nu\bar{p}_\psi} \nonumber\\
& + \frac{\gamma}{2\alpha}\sin^2\theta\cos^2\psi \left(\mu-2\nu\bar{p}_\psi
\right) - \epsilon\bar{p}_\psi \label{eqn:H}
\end{eqnarray}
and
\begin{eqnarray}
\fl
\bar{F}\left( \theta, \psi, \bar{p}_{\theta}, \bar{p}_{\psi}\right)
=\bar{p}_\psi + \frac{\nu}{\alpha}\cos\theta -
\frac{\sqrt{\mu-2\nu\bar{p}_\psi}}{\alpha}\left( \bar{p}_{\theta}\sin\psi -
\cos\psi \left( \frac{1 - \bar{p}_{\psi}\cos\theta}{\sin\theta} \right) \right).
\label{first_int}
\end{eqnarray}
We shall focus on the case $\alpha=O(1)$ and treat $\mu$, $\nu$, $\epsilon$
 and $\gamma$ as small parameters in a scaling to be made precise in the
next section (corresponding to a rod in a weak magnetic field with
extensibility/shearability as the perturbation).

In the absence of a magnetic field the parameter $\mu$ equals zero, provided
one (re)defines Euler angles relative to the constant force vector in the rod,
so that the polar singularity $\theta=0$ corresponds to the line of force.
(In the reduction mentioned above the Euler angles were quite naturally
defined relative to the direction of the magnetic field, which, unlike the
direction of force, is constant.) Both Hamiltonian and first integral then
reduce to the familiar expressions, the latter being equal to $\bar{p}_\psi$,
the moment about the force vector. A nonzero $\mu$ thus corresponds to a
different choice of coordinates taken inclined to the force vector; it has no
physical meaning. It will be used in the first perturbation in the next
section, however, to deflect the fixed point solution of the unperturbed
system away from the polar singularity so that a Melnikov analysis can be
applied. A similar inclination is used in \cite{K11} to unfold the fixed
point.

\section{Scaling of the equilibrium equations}

Dropping overbars,
we write the canonical equilibrium equations as
\begin{eqnarray}
\fl
&&
\dot{\theta}=p_\theta,\quad
\dot{\psi}=\frac{p_\psi-\cos\theta}{\sin^2\theta}
 -\frac{\nu(1+\gamma\cos\theta)\sin\theta\cos\psi}{\sqrt{\mu-2\nu p_\psi}}
 -\frac{\gamma\nu}{\alpha}\sin^2\theta\cos^2\psi-\epsilon,\nonumber\\
\fl
&&
\dot{p}_\theta=-\frac{(p_\psi-\cos\theta)(1-p_\psi\cos\theta)}{\sin^3\theta}
 +\alpha\sin\theta(1+\gamma\cos\theta)\nonumber\\[-1.5ex]
\fl
\label{eqn:em}\\[-1.5ex]
\fl
&& \qquad
-\left(\cos\theta+\gamma(\cos^2\theta-\sin^2\theta)\right)
\cos\psi\sqrt{\mu-2\nu p_\psi}
-\frac{\gamma}{\alpha}\sin\theta\cos\theta\cos^2\psi\,(\mu-2\nu p_\psi),
\nonumber\\
\fl
&&
\dot{p}_\psi=(1+\gamma\cos\theta)\sin\theta\sin\psi\sqrt{\mu-2\nu p_\psi}
 +\frac{\gamma}{\alpha}\sin^2\theta\cos\psi\sin\psi\,(\mu-2\nu p_\psi).\nonumber
\end{eqnarray}
If $\epsilon,\gamma=0$ (i.e., in the inextensible/unshearable limit), these equations
have the first integral $F$ given in (\ref{first_int}).

\subsection{The case $\epsilon=\gamma=\nu=\mu=0$}
Let $\epsilon=\gamma=\nu=\mu=0$.
Then the Hamiltonian in~\eref{eqn:H} becomes
\begin{equation}
H(\theta,\psi,p_\theta,p_\psi)
 =\frac{1}{2}p_\theta^2
 +\frac{1}{2}\left(\frac{p_\psi-\cos\theta}{\sin\theta}\right)^2+\alpha\cos\theta.
\label{eqn:H0}
\end{equation}
This is the standard reduced Hamiltonian for the isotropic Kirchhoff rod
\cite{HT00}. We easily see that the $(\theta,p_\theta)$-dynamics is
independent of $\psi$ and therefore $p_\psi$ is conserved.  
A Hamiltonian similar to \eref{eqn:H0} was discussed in Section~3 of
\cite{HM83}. In particular, it was shown that in the
$(\theta,p_\theta)$-dynamics there exists a hyperbolic saddle point at the
origin and a homoclinic orbit connecting it to itself at a single special
value of $p_\psi$. The saddle point corresponds to a periodic orbit
in the full four-dimensional phase space.
For \eref{eqn:H0} we have a similar situation.
If (and only if) $p_\psi=1$ and $\alpha>\case{1}{4}$ then in the
$(\theta,p_\theta)$-dynamics there exists a hyperbolic saddle at the origin
 and a pair of homoclinic orbits
 $(\theta,p_\theta)=(\pm\theta^\h(t),\pm p_\theta^\h(t))$, where
\begin{equation}
\theta^\h(t)=\pm\arccos\left(1-\frac{4\alpha-1}{2\alpha}
 \sech^2\!\left(\frac{\sqrt{4\alpha-1}}{2}t\right)\right),\quad
p_\theta^\h(t)=\dot{\theta}^\h(t).
\label{eqn:ho1}
\end{equation}
Along the homoclinic orbits we also estimate the variation of $\psi$
 from $0$ to $t$ as $\psi^\h(t)$, where
\begin{equation}
\psi^\h(t)=\int_{0}^t\frac{\d t}{1+\cos\theta^\h(t)}
 =\frac{t}{2}
 +\arctan\left(\sqrt{4\alpha-1}\tanh\left(\case{1}{2}\sqrt{4\alpha-1}\,t\right)
 \right).
\label{eqn:ho2}
\end{equation}

\subsection{The case $\epsilon=\gamma=\nu=0$ and $0<\mu\ll 1$}
Let $0<\mu\ll 1$ while $\epsilon=\gamma=\nu=0$.
Following the application of Melnikov theory to the nearly-symmetric heavy
top in \cite{HM83} one would seem to prove that there exists a hyperbolic
periodic orbit near $(\theta,p_\theta,p_\psi)=(0,0,1)$ which gets perturbed
to a periodic orbit with transverse intersections of its invariant manifolds.
However, this is not the case.
The problem is the polar singularity at $\theta=0$ as a result of which
the homoclinic orbit only exists for a single value of the action integral,
$F=p_\psi=1$. The 
 Melnikov result in \cite{HM83} requires the
homoclinic orbit to exist for an open interval of $F$ values so that
the partial derivative $\partial H/\partial F$ exists, 
as required to reduce the two-degrees-of-freedom Hamiltonian system
to a periodically perturbed planar system where the classical Melnikov
theorem \cite{M63,GH83} can be applied.

A direct analysis for \eref{eqn:em} with $\epsilon=\gamma=\nu=0$
 shows that there is not such a periodic orbit
 but instead two hyperbolic equilibria at
\begin{equation}
(\theta,\psi,p_\theta,p_\psi)
 =\left(
 \arctan\frac{\sqrt{\mu}}{\alpha},0,0,\frac{\alpha}{\sqrt{\alpha^2+\mu}}
\right)
\label{eqn:hypeq1}
\end{equation}
and
\begin{equation}
(\theta,\psi,p_\theta,p_\psi)
 =\left(
 -\arctan\frac{\sqrt{\mu}}{\alpha},\pi,0,\frac{\alpha}{\sqrt{\alpha^2+\mu}}
\right)
\label{eqn:hypeq2}
\end{equation}
near $(\theta,p_\theta,p_\psi)=(0,0,1)$.
We compute the Jacobian matrices
 for the right hand side of \eref{eqn:em} with $\epsilon=\gamma=\nu=0$
 at \eref{eqn:hypeq1} and \eref{eqn:hypeq2}, respectively, as
\[
\left(
\begin{array}{@{}cccc@{}}
0 & 0 & 1 & 0\\
\displaystyle\sqrt{\frac{\alpha^2+\mu}{\mu}} & 0 & 0
 & \displaystyle\frac{\alpha^2+\mu}{\mu}\\
-1+\sqrt{\alpha^2+\mu} & 0 & 0 & \displaystyle-\sqrt{\frac{\alpha^2+\mu}{\mu}}\\
0 & \displaystyle\frac{\mu}{\sqrt{\alpha^2+\mu}} & 0 &  0
\end{array}
\right)
\]
and 
\[
\left(
\begin{array}{@{}cccc@{}}
0 & 0 & 1 & 0\\
\displaystyle-\sqrt{\frac{\alpha^2+\mu}{\mu}} & 0 & 0
 & \displaystyle\frac{\alpha^2+\mu}{\mu}\\
-1+\sqrt{\alpha^2+\mu} & 0 & 0 & \displaystyle\sqrt{\frac{\alpha^2+\mu}{\mu}}\\
0 & \displaystyle\frac{\mu}{\sqrt{\alpha^2+\mu}} & 0 &  0
\end{array}
\right),
\]
both of which have eigenvalues
\[
\pm\sqrt{\sqrt{\alpha^2+\mu}-\case{1}{4}}\pm\case{1}{2}\,i.
\]
Hence, the equilibria \eref{eqn:hypeq1} and \eref{eqn:hypeq2} are hyperbolic saddles
 if $\alpha>\case{1}{4}$.
We easily see that they converge to $(\theta,\psi,p_\theta,p_\psi)=(0,0,0,1)$
 and $(0,\pi,0,1)$ as $\mu\to 0$.
Since the Hamiltonian system~\eref{eqn:em} is completely integrable,
 their stable and unstable manifolds cannot split.
Applying a standard asymptotic analysis
 which is rigorously based on Gronwall's inequality (see, e.g., \cite{M07}),
 we show that they have one-parameter families of homoclinic orbits given by
\begin{equation}
(\theta,\psi,p_\theta,p_\psi)
 =\left(\pm\theta^\h(t),\psi^\h(t)+\psi_0,\pm p_{\theta}^\h(t),1\right)
 +O(\sqrt{\mu})
\label{eqn:homo}
\end{equation}
for $t\in(-T_0,T_0)$ with $T_0=O(1)$,
 where $0\le\psi_0<2\pi$.

\subsection{The case $\epsilon=\gamma=0$ and $0<\nu\ll\mu\ll 1$}
Let $0<\nu\ll\mu$ while $\epsilon=\gamma=0$.
By persistence of hyperbolic equilibria (see Proposition~\ref{prop:A}),
 we see that there exist two hyperbolic equilibria
 in $O(\nu/\sqrt{\mu})$-neighbourhoods of \eref{eqn:hypeq1} and \eref{eqn:hypeq2}.
Since the Hamiltonian system~\eref{eqn:em} is still completely integrable,
 they have one-parameter families of homoclinic orbits
 in $O(\nu/\sqrt{\mu})$-neighbourhoods of \eref{eqn:homo}
 for $t\in(-T,T)$ with $T<T_0$ and $T=O(1)$.

\section{Melnikov-type analysis}
We now let $\gamma=\epsilon\hat{\gamma}$, where $\hat{\gamma}=O(1)$,
and assume that $0<\epsilon\ll\sqrt{\nu}\ll\sqrt{\mu}\ll 1$. We can then
apply the Melnikov-type technique of \cite{LU84,W88}, briefly reviewed in
Appendix~A, with 
\[
H_0=\frac{1}{2} p_{\theta}^{2} + \frac{1}{2}
\left( \frac{p_{\psi}-\cos\theta}{\sin\theta} \right)^{2} +\alpha\cos\theta
 +\sin\theta\cos\psi\sqrt{\mu-2\nu p_\psi}
\]
and
\[
\fl
H_1=-p_\psi + \frac{1}{2}\alpha\hat{\gamma}\cos^2\theta
 +\hat{\gamma}\sin\theta\cos\theta\cos\psi\sqrt{\mu-2\nu p_\psi}
 +\frac{\hat{\gamma}}{2\alpha}\sin^2\theta\cos^2\psi \left(\mu-2\nu p_\psi
\right).
\]
Note that hypotheses~(H1) and (H2) are satisfied with $K=F$ and
$\kappa=\psi_0$.

As stated in Proposition~\ref{prop:A},
 the two hyperbolic equilibria near \eref{eqn:hypeq1} and \eref{eqn:hypeq2}
 still persist but their stable and unstable manifolds may split.
Taking $T,T_0$ sufficiently large,
 we compute the Melnikov function \eref{eqn:M} as
\begin{eqnarray}
\fl
M(\psi_0)
&=&\int_{-\infty}^\infty\left(-\frac{\partial F}{\partial\psi}
 +\alpha\hat{\gamma}\frac{\partial F}{\partial p_\theta}
 \sin\theta(t)\cos\theta(t)\right.\nonumber\\
\fl && \qquad\left.+\hat{\gamma}\frac{\partial F}{\partial p_\psi}
 \sin\theta(t)\cos\theta(t)\sin\psi(t)\sqrt{\mu-2\nu p_\psi(t)}
 \right)\d t+O(\mu)\nonumber\\
\fl
&=&\frac{1}{\alpha}\int_{-\infty}^\infty
 \sqrt{\mu-2\nu p_\psi(t)}\left(p_\theta(t)\cos\psi(t)
 +\frac{1-p_\psi(t)\cos\theta(t)}{\sin\theta(t)}\sin\psi(t)\right)\d t+O(\mu)
\nonumber\\
\fl
&=&\pm\frac{\sqrt{\mu-2\nu}\,\Delta(\alpha)}{\alpha}\sin\psi_0+O(\mu),
\label{eqn:dI}
\end{eqnarray}
where the above integrands are evaluated on the homoclinic orbits
 for $\mu,\nu\neq 0$ and
\begin{equation}
\Delta(\alpha)=\int_{-\infty}^\infty
 \left[\frac{\sin\theta^\h(t)}{1+\cos\theta^\h(t)}
 \cos\psi^\h(t)-p_\theta^\h(t)\sin\psi^\h(t)\right]\d t.
\label{eqn:dI0}
\end{equation}
Here we have used the fact that $\theta^\h(t)$ is an even function of $t$,
 $p_\theta^\h(t),\psi^\h(t)$ are odd functions
 and the integrand of \eref{eqn:dI} tends to zero exponentially
 as $t\to\pm\infty$.
Note that $M(\psi_0)$ is independent of $\hat{\gamma}$ up to $O(\sqrt{\mu})$.

\begin{figure}
\begin{center}
\includegraphics[scale=0.6]{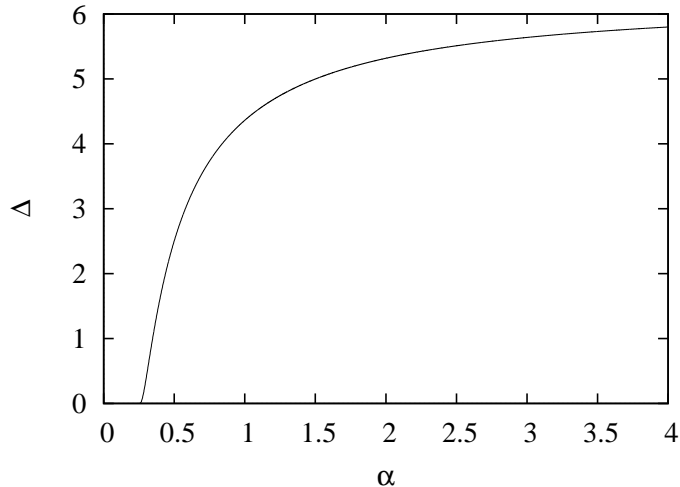}
\end{center}
\caption{Plot of the function $\Delta(\alpha)$.
\label{fig:int}}
\end{figure}

Surprisingly, we can compute the integral \eref{eqn:dI0} analytically as follows.
Let $\delta=\sqrt{\alpha-\case{1}{4}}$.
By \eref{eqn:ho1} and \eref{eqn:ho2}, we have
\begin{eqnarray*}
&&
\frac{\sin\theta^\h(t)}{1+\cos\theta^\h(t)}
 =\pm\frac{2\delta\sech\,\delta t}{\sqrt{4\delta^2\tanh^2\!\delta t+1}},\quad
p_\theta^\h(t)
 =\mp\frac{4\sech\,\delta t\tanh\delta t}{\sqrt{4\delta^2\tanh^2\!\delta t+1}},\\
&&
\psi^\h(t)
 =\case{1}{2}t+\arctan(2\delta\tanh\delta t).
\end{eqnarray*}
Using the above relations and noting that
\begin{eqnarray*}
&&
\cos\left(\arctan(2\delta\tanh\delta t)\right)
=\frac{1}{\sqrt{4\delta^2\tanh^2\!\delta t+1}},\\
&&
\sin\left(\arctan(2\delta\tanh\delta t)\right)
=\frac{2\delta\tanh\delta t}{\sqrt{4\delta^2\tanh^2\!\delta t+1}}
\end{eqnarray*}
we can rewrite the integral \eref{eqn:dI0} as
\[
2\delta\int_{-\infty}^{\infty}\sech\,\delta t\cos\case{1}{2}t\,\d t
=2\pi\sech\left(\frac{\pi}{4\delta}\right),
\]
i.e.,
\begin{equation}
\Delta(\alpha)=2\pi\sech\left(\frac{\pi}{2\sqrt{4\alpha-1}}\right)\neq 0
\label{eqn:Delta}
\end{equation}
for $\alpha>\case{1}{4}$.
The function $\Delta(\alpha)$ is plotted in \fref{fig:int}.

Thus, if $\epsilon\neq 0$ the Melnikov function $M(\psi_0)$ has a simple zero
near $\psi_0=0,\pi$ and by Theorem~\ref{thm:A} there exist transverse
homoclinic orbits to equilibria near \eref{eqn:hypeq1} and \eref{eqn:hypeq2}.
This result holds whether or not $\gamma=0$.
Moreover, it follows from Devaney's theorem \cite{D76}
 that the system is chaotic in the neighbourhood of the homoclinic orbit
 in the sense that any local transverse section to the homoclinic orbit
 contains a compact, invariant, hyperbolic set on which the Poincar\'e map
 is topologically conjugate to a Bernoulli shift of finite type. This in turn implies that the Hamiltonian $H$
has no analytic conserved quantities independent of $H$ itself, i.e., the
corresponding Hamiltonian system is nonintegrable \cite{MH88}.

\section{Numerical analysis}

Here we give numerical evidence of the above theoretical results.
We first describe our approach to compute the stable and unstable manifolds.

We write \eref{eqn:em} as
\begin{equation}
\dot{\xi}=J\D H(\xi),
\label{eqn:em1}
\end{equation}
where $\xi=(\theta,\psi,p_\theta,p_\psi)$
 and $J$ is the $4\times 4$ symplectic matrix,
\[
J=\left(
\begin{array}{@{}cc@{}}
0 & \id_2\\
-\id_2 & 0
\end{array}
\right),
\]
with $\id_2$ the $2\times 2$ identity matrix.
Let $\xi_0$ denote the equilibria near \eref{eqn:hypeq1} and \eref{eqn:hypeq2}.
We compute $W^\u(\xi_0)$ as follows.

We approximate $W^\u(\xi_0)$ by the unstable subspace of the linearised system
\begin{equation}
\dot{\eta}=J\D^2 H(\xi_0)\eta
\label{eqn:lem}
\end{equation}
near $\xi=\xi_0$.
Under this approximation, we obtain an orbit $\xi(t)$ on $W^\u(\xi_0)$
 as a solution of $\eref{eqn:em1}$ satisfying the boundary conditions
\begin{equation}
L_\s\xi(0)=0,\quad
\xi(T_\u)=\xi_0^\u,
\label{eqn:bcu}
\end{equation}
where $T_\u>0$ is a constant, $L_\s$ is a $2\times 4$ matrix consisting
 of row eigenvectors corresponding to eigenvalues with negative real parts
 of the Jacobian matrix $J\D^2 H(\xi_0)$,
 and $\xi_0^\u$ is a point on $W^\u(\xi_0)$.
Thus, we solve the boundary value problem \eref{eqn:em1} and \eref{eqn:bcu}
 and continue the solution in $\xi_0^\u$ to compute $W^\u(\xi_0)$ numerically.

\begin{figure}[t]
\begin{center}
\includegraphics[scale=0.6]{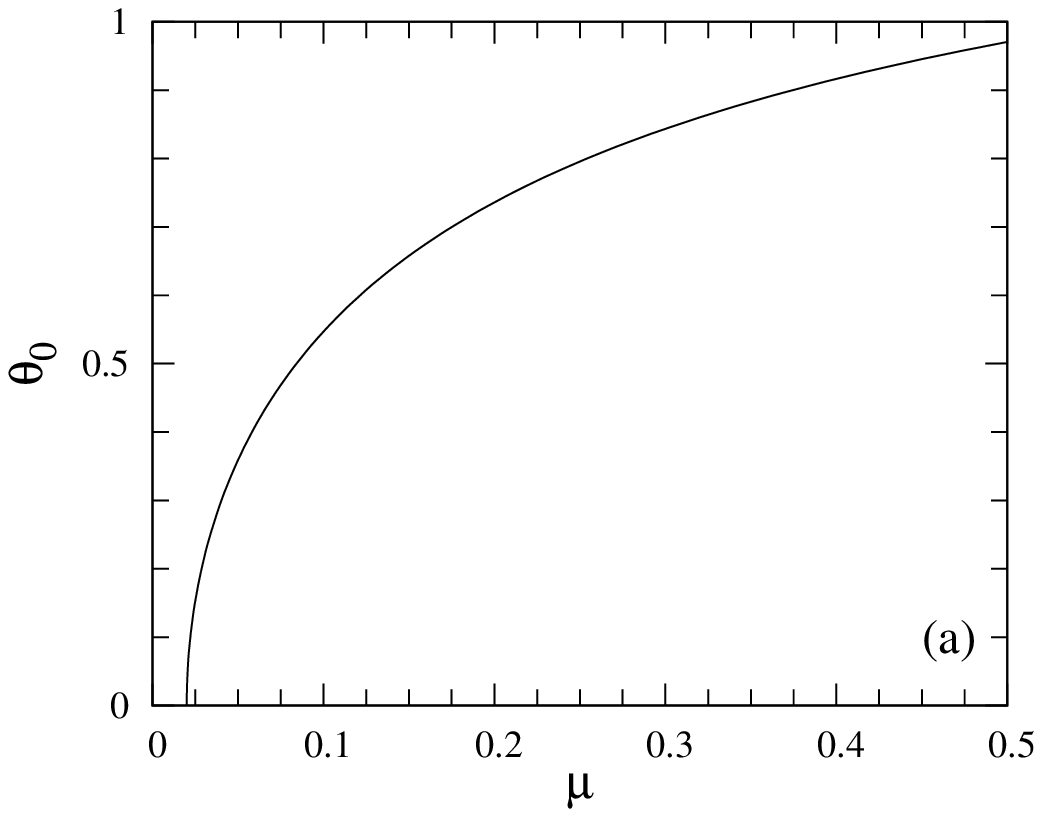}\quad
\includegraphics[scale=0.6]{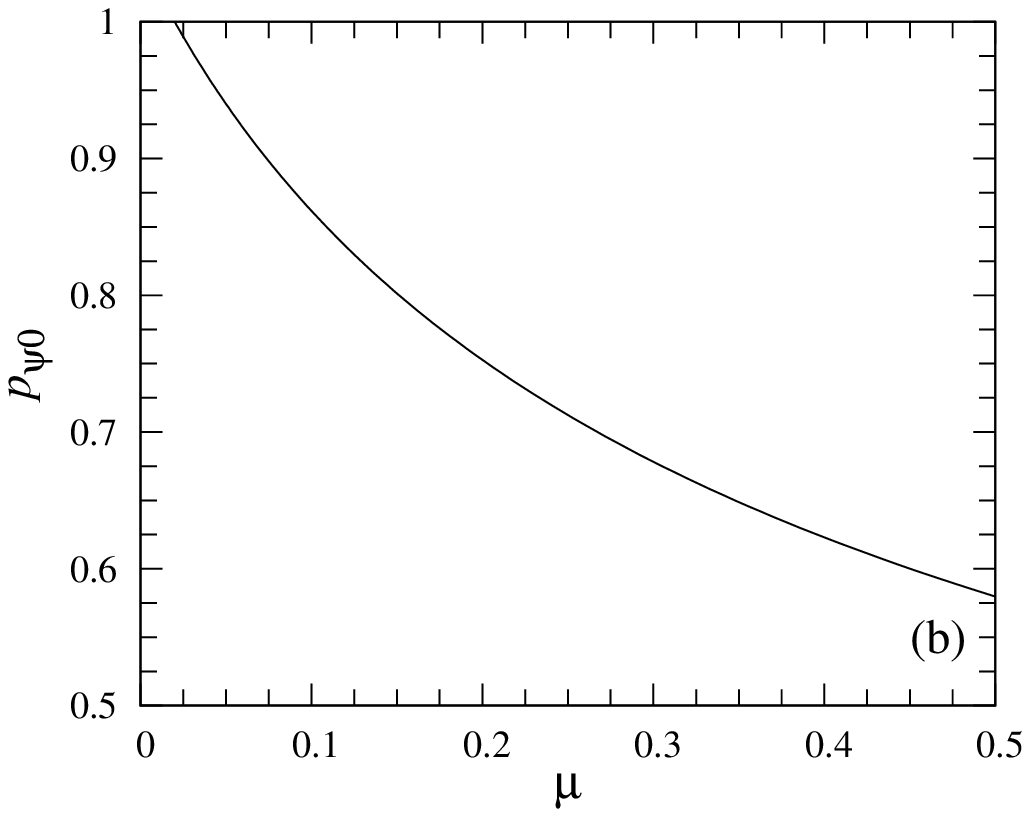}
\end{center}
\caption{Numerically computed equilibria in \eref{eqn:em}
 for $\epsilon=\nu=0.01$, $\gamma=0$ and $\alpha=0.5$.
\label{fig:eq}}
\end{figure}

Similarly, to compute $W^\s(\xi_0)$,
 we solve the boundary value problem \eref{eqn:em1} and
\begin{equation}
L_\u\xi(0)=0,\quad
\xi(-T_\s)=\xi_0^\s,
\label{eqn:bcs}
\end{equation}
where $T_\s>0$ is a constant, $L_\u$ is a $2\times 4$ matrix consisting
 of row eigenvectors corresponding to eigenvalues with positive real parts
 of the Jacobian matrix $J\D^2 H(\xi_0)$,
 and $\xi_0^\s$ is a point on $W^\s(\xi_0)$.
The intersection of $W^\s(\xi_0)$ and $W^\u(\xi_0)$
 gives a homoclinic orbit to $\xi_0$. 
We adopt a solution of the linearised system \eref{eqn:lem} with $T_\u,T_\s$
small
 as a starting solution,
 and use the numerical continuation tool AUTO97 \cite{AUTO97}
 to perform the above computations.
In these computations
 we can also take $T_{\s,\u}$, $\xi(0)$ and $\xi_0^{\u,\s}$ as free parameters.

\Fref{fig:eq} shows a numerically computed equilibrium $(\theta_0,p_{\psi 0})$
 near \eref{eqn:hypeq1} in \eref{eqn:em} for $\epsilon=\nu=0.01$, $\gamma=0$
 and $\alpha=0.5$.
Note that $(\psi,p_\theta)=(0,0)$ at the equilibrium
 and there is another equilibrium
 at $(\theta,\psi,p_\theta,p_\psi)=(-\theta_0,\pi,0,p_{\psi 0})$
 near \eref{eqn:hypeq2}.
These equilibria exist only for $\mu>2\nu=0.02$.

\begin{figure}[t]
\begin{center}
\includegraphics[scale=0.6]{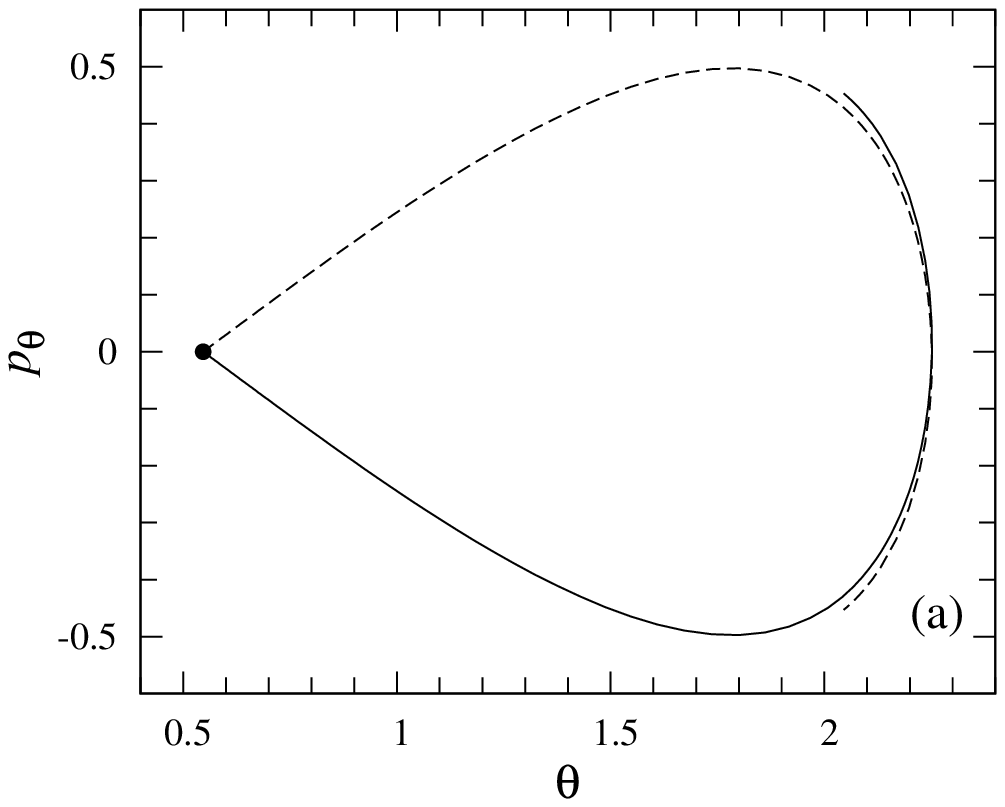}\quad
\includegraphics[scale=0.6]{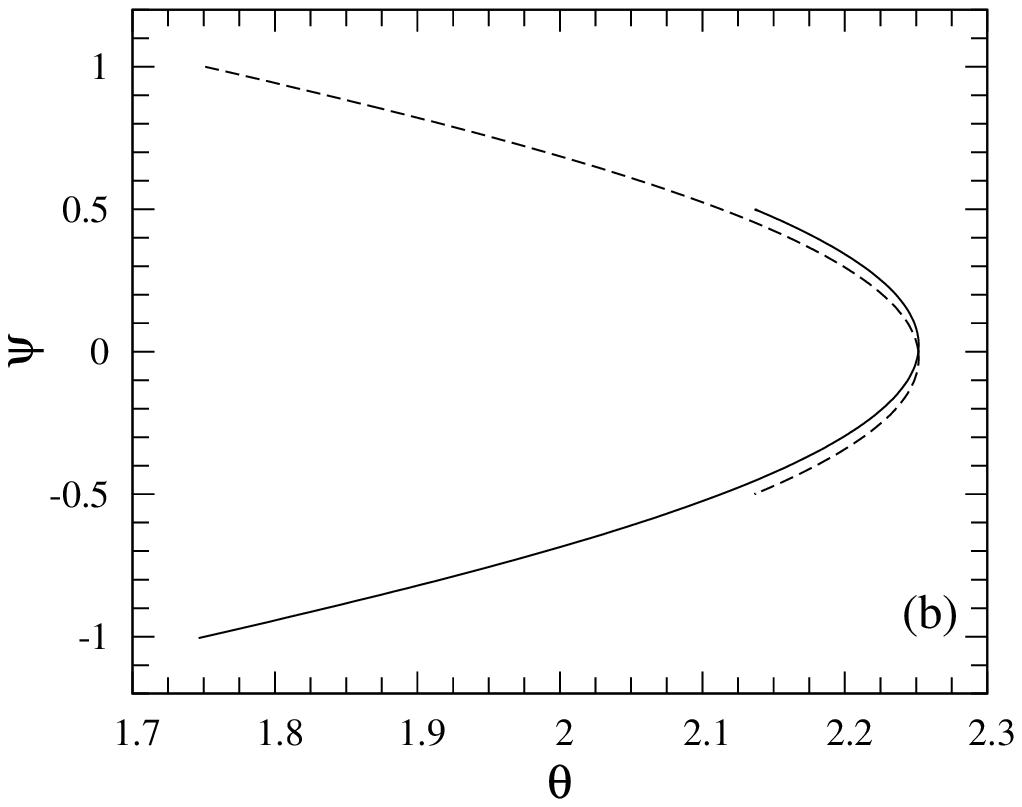}
\caption{Numerically computed stable and unstable manifolds of the equilibrium
 near $(\theta,\psi,p_\theta,p_\psi)=(\theta_0,0,0,p_{\psi 0})$ in \eref{eqn:em}
 for $\mu=0.1$, $\epsilon=\nu=0.01$, $\gamma=0$ and $\alpha=0.5$:
(a) $\psi=0$; (b) $p_\theta=0$.
The stable and unstable manifolds
 are plotted as solid and broken lines, respectively.
In plate~(a) ``$\bullet$'' represents the position of the equilibrium.
\label{fig:m0.1}}
\end{center}
\end{figure}

\begin{figure}[t]
\begin{center}
\includegraphics[scale=0.6]{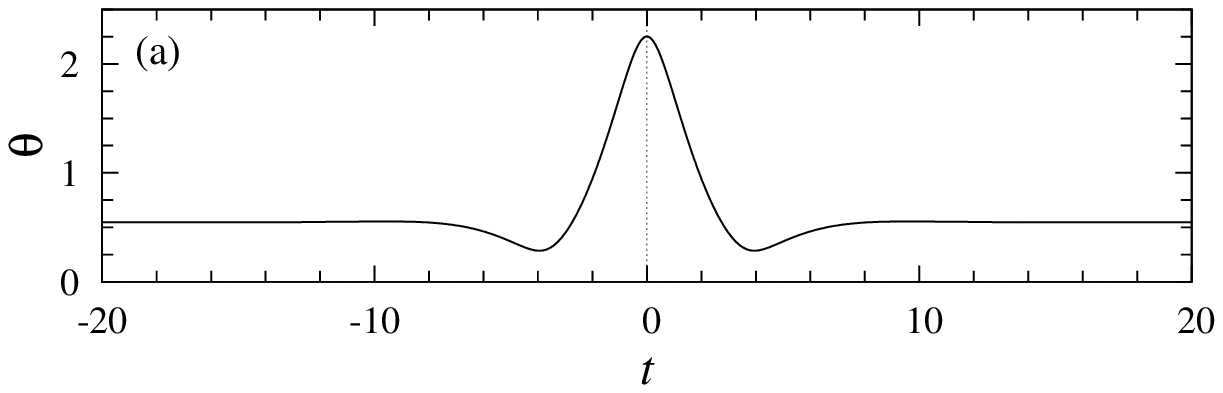}\quad
\includegraphics[scale=0.6]{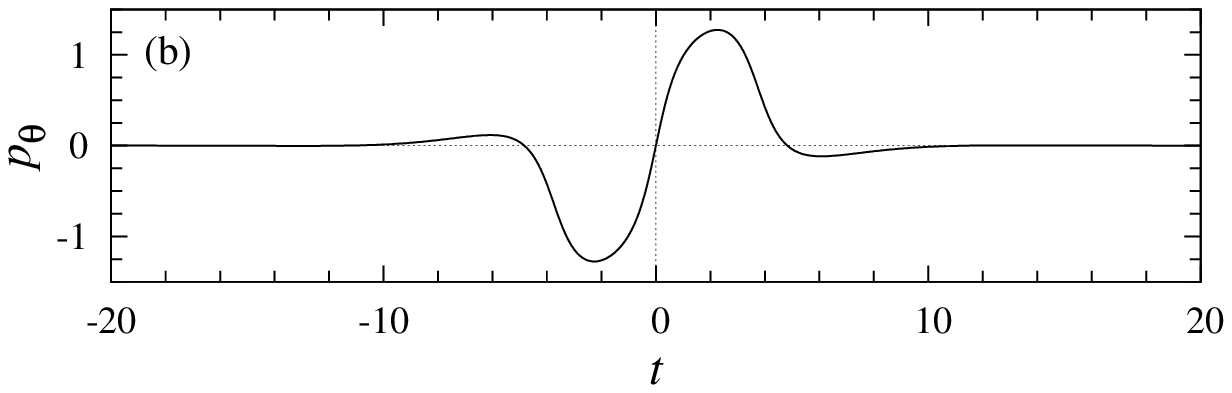}\\[2ex]
\includegraphics[scale=0.6]{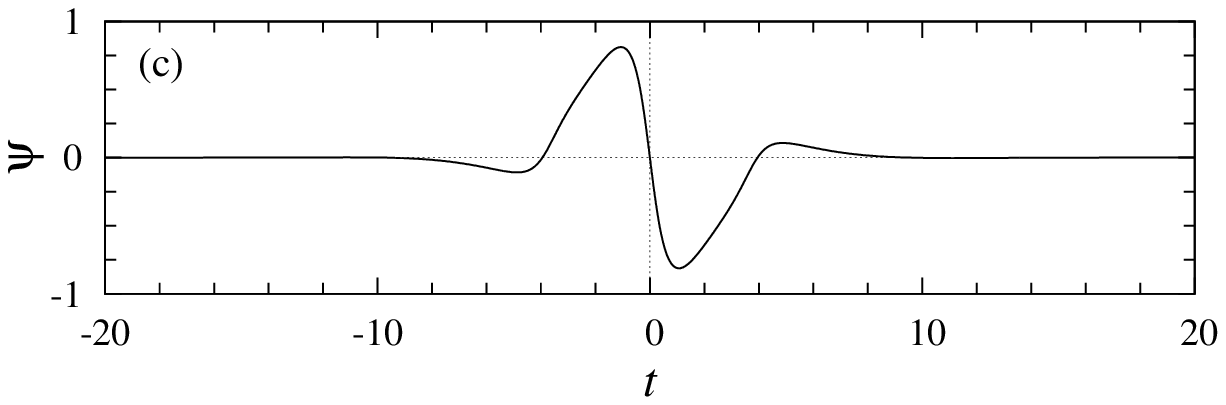}\quad
\includegraphics[scale=0.6]{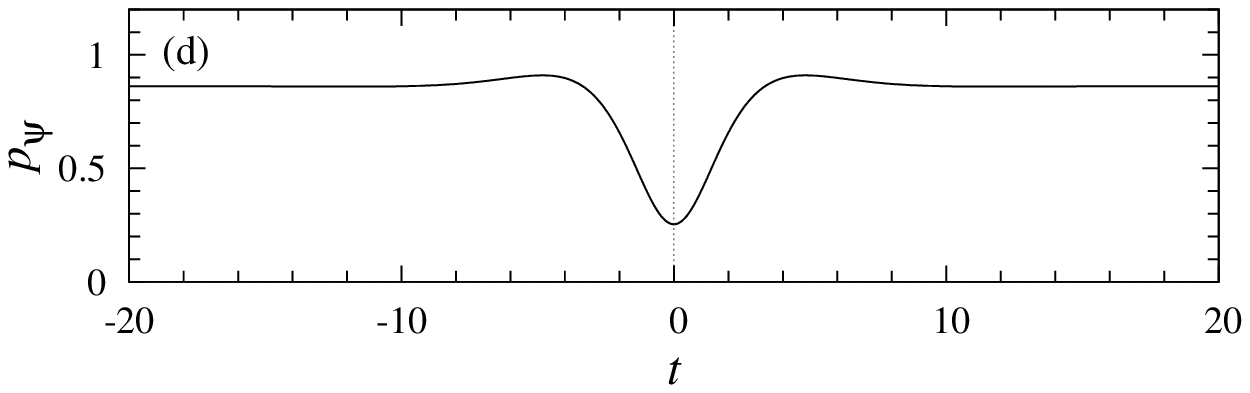}
\caption{Numerically computed transverse homoclinic orbit in \eref{eqn:em}
 for $\mu=0.1$, $\epsilon=\nu=0.01$, $\gamma=0$ and $\alpha=0.5$.
\label{fig:o0.1}}
\end{center}
\end{figure}

\Fref{fig:m0.1} shows $\{\psi=0\}$ and $\{p_\theta=0\}$ sections of
numerically computed stable and unstable manifolds of the equilibrium
$(\theta,\psi,p_\theta,p_\psi)=(\theta_0,0,\psi,p_{\psi 0})$ near
\eref{eqn:hypeq1} in \eref{eqn:em} for $\mu=0.1$, $\epsilon=\nu=0.01$,
 $\gamma=0$ and $\alpha=0.5$.
From these sections
 we see that these manifolds intersect transversely in the energy level set
 and there exists a transverse homoclinic orbit, as predicted by the theory
 in Section~4.
We note that
 other branches of these manifolds, which intersect $\{\theta=0\}$,
 are very difficult to compute
 due to the singularity in \eref{eqn:em}.
The transverse homoclinic orbit detected in \fref{fig:m0.1}
 is plotted in \fref{fig:o0.1}.

\section{Conclusion}

We have shown that the reduced equilibrium equations for an extensible and
shearable conducting rod in a uniform magnetic field possess transverse
homoclinic orbits to saddle-type equilibria. Via a classical result in
dynamical systems this implies that the equations are nonintegrable and
give rise to chaotic dynamics.
To obtain the result, we applied a version of Melnikov's method,
 taking the classical Kirchhoff rod as the unperturbed problem
 and introducing magnetic and elastic effects as successive perturbations.

Our method is also of interest because of the way it deals with the
Euler-angle singularity of the reduced equations. Such a singularity is
quite common in mechanical problems, e.g., in rigid-body dynamics.
Indeed, it was present in the nearly-symmetric heavy top example
treated in \cite{HM83}. As pointed out elsewhere \cite{K11}, in this
example the conditions required for the Melnikov method
are not satisfied and the application is therefore invalid, although there
can be no doubt that the result is true (i.e., the nearly-symmetric heavy
top is nonintegrable). In a certain limit, namely $\gamma=\nu=\epsilon=0$,
$\mu\ll\alpha=O(1)$, our magnetic rod gives an example where a similar
cavalier application of the Melnikov method would suggest that the system
is not integrable, yet the presence of the explicit first integral $F$
implies that the system is in fact integrable. The resolution is that the
fixed point ($\theta$,$p_\theta)=(0,0)$ of the planar unperturbed system is
not perturbed into a periodic orbit of the $\mu$-perturbed system. Rather,
two fixed points appear (cf.~(\ref{eqn:hypeq1}) and (\ref{eqn:hypeq2})),
unfolding the Euler-angle singularity. As explained in Section 2,
the parameter $\mu$ is an artifical parameter if no magnetic field is
present. Thus we see that in our Melnikov application the singularity is
circumvented by the introduction of an artificial unfolding
that does not affect the physics of the problem. A similar technique (an
inclined plane) is used in \cite{K11} to give a correct proof of the
nonintegrability of the nearly-symmetric heavy top. The method may be
applicable to other systems.

\ack
KY appreciates support from the JSPS
through Grant-in-Aid for Scientific Research (C) Nos.~21540124 and 22540180.

\appendix

\newtheorem{theorem}{Theorem}[section]
\newtheorem{proposition}[theorem]{Proposition}
\newtheorem{lemma}[theorem]{Lemma}
\newtheorem{remark}[theorem]{Remark}

\section{Melnikov-type technique}

\renewcommand{\thesection}{\Alph{section}}

For the reader's convenience,
 we briefly review an analytical technique to detect
 the existence of transverse homoclinic orbits to hyperbolic saddles
 in two-degrees-of-freedom, nearly-integrable Hamiltonian systems.
See \cite{LU84,W88} for details.

Consider a two-degrees-of-freedom Hamiltonian systems of the form
\begin{equation}
\dot{x}=J\D H_0(x)+\epsilon J\D H_1(x),\quad
x\in\Rset^4,
\label{eqn:Hsys}
\end{equation}
where $H_0,H_1:\Rset^4\rightarrow\Rset$ are $C^{r+1}$ ($r\ge 2$).
When $\epsilon=0$,
 equation~\eref{eqn:Hsys} becomes
\begin{equation}
\dot{x}=J\D H_0(x).
\label{eqn:Hsys0}
\end{equation}
We make the following assumptions on \eref{eqn:Hsys0}.
\begin{list}{}{
\setlength{\leftmargin}{3em}
\setlength{\labelwidth}{3em}
}
\item[\bf(H1)]
The system~\eref{eqn:Hsys0} has a first integral $K(x)$
 such that $H_0(x)$ and $K(x)$ are independent,
 i.e., $\D H_0(x)$ and $\D K(x)$ are linearly independent, 
 and in involution, i.e., their Poisson bracket is zero,
\begin{equation}
\{K,H_0\}\equiv\D K(x)\cdot J\D H_0(x)=0.
\label{eqn:inv}
\end{equation}
In other words, the system~\eref{eqn:Hsys0} is completely integrable.
\item[\bf(H2)]
There is a hyperbolic saddle point at $x=x_0$
 possessing a one-parameter family of homoclinic orbits $x^\h(t;\kappa)$,
 $\kappa\in\I$, where $\I\subset\Rset$ is a nonempty open interval.
In addition, $x^\h(t;\kappa)$ is $C^r$ not only in $t$ but also in $\kappa$.
\end{list}
The Hamiltonian system~\eref{eqn:Hsys} also
 has the System~III form of \cite{W88} with $n=2$ and $m=0$.
 
Since equation~\eref{eqn:Hsys0} is Hamiltonian and four-dimensional,
 the hyperbolic saddle $x=x_0$ has two-dimensional stable and unstable manifolds,
 which are denoted by $W^{\s,\u}(x_0)$.
Let $\Gamma=\{x=x^\h(t;\kappa)\,|\,t\in\Rset,\kappa\in\I\}\cup\{x_0\}$.
It follows from assumption~(H2) that
\[
W^\s(x_0)\cap W^\u(x_0)\supset\Gamma.
\]
By a rather standard result on persistence of hyperbolic equilibria
 and their stable and unstable manifolds (see, e.g., \cite{M07}),
 we immediately have the following result. 

\begin{proposition}
\label{prop:A}
For $\epsilon>0$ sufficiently small
 the system~\eref{eqn:Hsys} has a hyperbolic equilibrium
 at $x=x_\epsilon=x_0+\O(\epsilon)$.
Moreover, the stable and unstable manifolds
 $W^{\s,\u}(x_\epsilon)$ of $x_\epsilon$
 are $\O(\epsilon)$-close to $W^{\s,\u}(x_0)$.
\end{proposition}

Now define the Melnikov function as
\begin{equation}
M(\kappa)
 =\int_{-\infty}^\infty\D K(x_0^\h(t;\kappa))\cdot
 J\D H_1(x_0^\h(t;\kappa))\d t.
\label{eqn:M}
\end{equation}
We have the following result, as shown in \cite{LU84}.

\begin{theorem}
\label{thm:A}
Suppose that $M(\kappa)$ has a simple zero at $\kappa=\kappa_0$, i.e.,
\[
M(\kappa_0)=0,\quad
\frac{\d M}{\d\kappa}(\kappa_0)\neq 0.
\]
Then for $\epsilon>0$ sufficiently small
 the stable and unstable manifolds $W^{\s,\u}(x_\epsilon)$
 intersect transversely and there exists a transverse homoclinic orbit
 $x=x^\h(t;\kappa_0)+\O(\epsilon)$ to the hyperbolic saddle $x_\epsilon$.
\end{theorem}

\end{document}